\documentclass[twocolumn,showpacs,preprintnumbers,prl]{revtex4}
\usepackage{amssymb}
\usepackage{amsmath}
\usepackage{graphicx}
\usepackage{dcolumn}
\usepackage{bm}

\begin{document}

\title{Generalized Bell inequality for mixed states with variable constraints%
}
\author{Chang-shui Yu}
\author{He-shan Song}
\email{hssong@dlut.edu.cn}
\affiliation{Department of Physics, Dalian University of Technology, Dalian 116024, P. R.
China}
\date{\today }

\begin{abstract}
In this paper, we present a generalized Bell inequality for mixed states.
The distinct characteristic is that the inequality has variable bound
depending on the decomposition of the density matrix. The inequality has
been shown to be more refined than the previous Bell inequality. It is
possible that a separable mixed state can violate the Bell inequality.
\end{abstract}

\maketitle

The concept of local realism is that physical systems may be described by
local objective properties that are independent of observation [1,2]. Bell
established that quantum theory is incompatible with local realism by
analyzing the special case of two spin-1/2 particles coupled in an angular
momentum singlet state [3]. In particular, the constraints on the statistics
of physically separated systems, called Bell inequality that can be violated
by the statistical predictions of quantum mechanics, is implied. In general,
the Bell inequality can be written as a locally realistic bound $\beta _{LR}$
on the expectation value of some Hermitian operator $\mathcal{\hat{B}}$
(Bell operator), i.e. $\left\langle \mathcal{\hat{B}}\right\rangle \leq
\beta _{LR}$ [2]. However, it is not all the entangled states that violate
the conventional Bell inequality [3,4,5]. In fact, if it is considered
quantum nonlocal, it is not necessary for a state to violate all possible
Bell's inequalities, as implied in Ref. [6-8]. The violation of any Bell's
inequality can show a given state to be nonlocal. Therefore, the uncovery of
quantum locality depends not only on the given quantum state but also on the
\textquotedblleft Bell operator\textquotedblright . That is to say, in order
to uncover the quantum locality of a given quantum state, one must construct
a proper Bell inequality or Bell operator.

Since the original Bell inequality was introduced [3] and developed
by Clauser, Horne, Shimony and Holt (CHSH) [4], the investigation of
Bell inequality has attracted a lot of attentions [9-12]. However,
only\ the case of pure states is completely solved [3,4,9,10], for
density matrices i.e. mixed states, only partial results have been
obtained so far [11-12]. In this paper, we present a generalized
Bell inequality for mixed states. The distinct characteristic is
that the inequality has variable bound depending on the
decomposition of the density matrix, i.e. the concrete realization
of the density matrix. By the study of Werner states [13] and
maximally entangled mixed states [14], the inequality has been shown
to be more refined than the previous Bell inequalities. We also show
a surprising result that a separable state may violate the Bell
inequality. Even though a potential understanding of the violation
for a separable mixed state has been provided finally, a deeper one
remains open.

At first, we will follow the analogous procedure to Ref. [4] to give our
Bell inequality.

Suppose we have an ensemble of particle pairs with $\varrho $ the density
matrix. We measure $A(a,\lambda )$ and $B(b,\lambda )$ on the two particles
of each pair, respectively, with $\left\vert A(a,\lambda )\right\vert \leq 1$
and $\left\vert B(b,\lambda )\right\vert \leq 1$. In particular, note that $%
a $ and $b$ are adjustable apparatus parameters and $\lambda $ is the hidden
variables with the normalized probability distribution $\rho (\lambda )$ for
the given quantum mechanical state. Furthermore, $A(a,\lambda )$ independent
of $b$ and $B(b,\lambda )$ independent of $a$ are required due to the
locality. All above are analogous to Ref. [4].

Defining the correlation function $P(a,b)=\int_{\Gamma }A(a,\lambda
)B(b,\lambda )\rho (\lambda )d\lambda $, where $\Gamma $ is the total $%
\lambda $ space, we have%
\begin{eqnarray}
&&\left\vert P(a,b)-P(a,c)\right\vert   \notag \\
&=&\left\vert \int_{\Gamma }\left[ A(a,\lambda )B(b,\lambda )-A(a,\lambda
)B(c,\lambda )\right] \rho (\lambda )d\lambda \right\vert .
\end{eqnarray}%
$\Gamma $ can always be divided into different regions denoted by $\Gamma
_{i}$ with
\begin{equation*}
\sum_{i=1}^{N}\int_{\Gamma _{i}}\rho \left( \lambda \right) d\lambda
=1,i=1,2,\cdot \cdot \cdot ,N,
\end{equation*}%
where $N$ represents the number of different regions. In the different
regions, there may be different correlations. Therefore, eq. (1) can be
rewritten analogous to Ref. [4] by
\begin{eqnarray}
&&\left\vert P(a,b)-P(a,c)\right\vert   \notag \\
&=&\sum_{j=1}^{n}\left( \int_{\Gamma _{j}}A(a,\lambda )B(b,\lambda )\left[
1\pm A(d,\lambda )B(c,\lambda )\right] \rho (\lambda )d\lambda \right.
\notag \\
&&\left. \left. -\int_{\Gamma _{j}}A(a,\lambda )B(c,\lambda )\left[ 1\pm
A(d,\lambda )B(b,\lambda )\right] \rho (\lambda )d\lambda \right)
\right\vert   \notag \\
&&+\left\vert \sum_{i=n+1}^{N}\int_{\Gamma _{i}}\left[ A(a,\lambda
)B(b,\lambda )-A(a,\lambda )B(c,\lambda )\right] \rho (\lambda )d\lambda
\right. ,
\end{eqnarray}%
where $n=1,2,\cdot \cdot \cdot ,N$. According to the inequality $\left\vert
\int f(x)dx\right\vert \leq \int \left\vert f(x)\right\vert dx$, one can
obtain
\begin{eqnarray}
&&\left\vert P(a,b)-P(a,c)\right\vert   \notag \\
&\leq &\sum_{j=1}^{n}\left( \int_{\Gamma _{j}}\left\vert 1\pm A(d,\lambda
)B(c,\lambda )\right\vert \rho (\lambda )d\lambda \right.   \notag \\
&&+\left. \int_{\Gamma _{j}}\left\vert 1\pm A(d,\lambda )B(b,\lambda
)\right\vert \rho (\lambda )d\lambda \right)   \notag \\
&&+\left\vert \sum_{i=n+1}^{N}\int_{\Gamma _{i}}\left[ A(a,\lambda
)B(b,\lambda )-A(a,\lambda )B(c,\lambda )\right] \rho (\lambda )d\lambda
\right\vert ,
\end{eqnarray}%
\bigskip where $\left\vert A(x^{\prime },\lambda )B(y^{\prime },\lambda
)\right\vert \leq 1$ with $x^{\prime },y^{\prime }=a,b,c$ or $d$ is implied.
It is obvious that if $N=1$, eq. (3) will reduce to the original CHSH
inequality [4] for pure states.

Consider the density matrix $\varrho =\sum_{i}p_{i}\left\vert \Phi
_{i}\right\rangle \left\langle \Phi _{i}\right\vert $, the correlation
function $P(a,b)$ can always be expressed by the joint measurement of
observables $\mathcal{A}(a)$ and $\mathcal{B}(b)$. That is to say, $P(a,b)$
can be obtained by
\begin{eqnarray}
P(a,b) &=&tr\varrho \left[ \mathcal{A}(a)\otimes \mathcal{B}(b)\right]
=\left\langle ab\right\rangle  \notag \\
&=&\sum_{i}p_{i}\left\langle \Phi _{i}\right\vert \mathcal{A}(a)\otimes
\mathcal{B}(b)\left\vert \Phi _{i}\right\rangle =\sum_{i}p_{i}\left\langle
ab\right\rangle _{i},
\end{eqnarray}%
where $\left\langle ab\right\rangle _{i}=\left\langle \Phi _{i}\right\vert
\mathcal{A}(a)\otimes \mathcal{B}(b)\left\vert \Phi _{i}\right\rangle $. If $%
\Gamma $ is divided as mentioned above into regions which just correspond to
the given decomposition $\left\{ p_{i},\Phi _{i}\right\} $ such that
\begin{equation}
\int_{\Gamma _{i}}A(a,\lambda )B(b,\lambda )\rho (\lambda )d\lambda
=p_{i}\left\langle ab\right\rangle _{i},
\end{equation}%
eq. (3) can be rewritten based on the expectation values of the observables
as%
\begin{equation}
\left\vert \left\langle ab\right\rangle -\left\langle ac\right\rangle
\right\vert \leq 2-\sum_{i=1}^{N}p_{i}\left\vert \left\langle
db\right\rangle _{i}+\left\langle dc\right\rangle _{i}\right\vert ,
\end{equation}%
when $n=N$ [16]. In fact, the inequality (6) will have different forms for
different $n$. Compared with the original CHSH inequality, the most
difference lies in that the expectation value of Bell operator [2], i.e. $%
\left\langle ab\right\rangle -\left\langle ac\right\rangle $ in inequality
(6), is constrained by a variable bound which depends on not only the
decomposition of $\varrho $ but also the apparatus parameters $b$, $c$ and $%
d $.

As a special case, if let $d=c$ and the $\Gamma _{j}$ parts are perfect
correlated, the inequality (6) can also be converted to
\begin{eqnarray}
\left\vert \left\langle ab\right\rangle -\left\langle ac\right\rangle
\right\vert &\leq &\left\vert \sum_{i=n+1}^{N}p_{i}\left( \left\langle
ab\right\rangle _{i}-\left\langle ac\right\rangle _{i}\right) \right\vert
\pm _{\Gamma _{j}}\sum_{j=1}^{n}p_{j}\left\langle \left\langle
bc\right\rangle \right\rangle _{j},  \notag \\
n &=&1,2,\cdot \cdot \cdot ,N
\end{eqnarray}%
where
\begin{equation}
\left\langle \left\langle bc\right\rangle \right\rangle _{j}=\left\langle
bb\right\rangle _{j}-\left\langle bc\right\rangle _{j}
\end{equation}%
and $"\pm _{\Gamma _{j}}"=\left\{
\begin{array}{cc}
"+", & \left\langle \left\langle bc\right\rangle \right\rangle >0 \\
"-", & \left\langle \left\langle bc\right\rangle \right\rangle \leq 0%
\end{array}%
\right. $. This is a generalization of the original Bell inequality [3].
Obviously, the inequality will reduce to the original Bell inequality if $%
N=1 $.

Consider a composite quantum system described in a Hilbert space $\mathcal{H}%
=\mathcal{H}^{1}\otimes \mathcal{H}^{2}$. The corresponding density matrix
can be given by $\varrho $, i.e. an operator with $\varrho =\varrho
^{\dagger }$, $tr\varrho =1$ and $\varrho \geq 0$. The density is separable
or classically correlated if there exists a decomposition $\left\{
p_{i},\varphi _{i}\otimes \psi _{i}\right\} $ such that $\varrho
=\sum_{i}p_{i}\left\vert \varphi _{i}\right\rangle \left\langle \varphi
_{i}\right\vert \otimes \left\vert \psi _{i}\right\rangle \left\langle \psi
_{i}\right\vert $. Otherwise, the density matrix is inseparable or EPR
correlated. Since our inequalities given by eq. (6) are derived from a local
hidden variable model, the joint measurements on the separable density
matrix $\varrho $ should be constrained by our variable bound for any $%
n=1,2,\cdot \cdot \cdot $. In other words, the violation of the inequality
(6) or (7) implies the existence of EPR correlation. Recalling the previous
Bell inequalities [3,4], the violation of the inequalities for mixed states
usually becomes more difficult than that for pure states. That is to say,
not all entangled states can be demonstrated to violate the inequalities.
The most familiar examples should be the Werner states [5] and the maximally
entangled mixed states [12]. However, the states (defined in $\left( 2\times
2\right) -$dimensional Hilbert space) will be shown to violate the
inequality given here. In this sense, we say that the inequality with
current form seems to be more refined than the previous ones [3,4].

The maximally entangled mixed state predicted by White et al. [14] has the
explicit form
\begin{equation}
\varrho _{m}=\left(
\begin{array}{cccc}
g(\gamma ) & 0 & 0 & \frac{\gamma }{2} \\
0 & 1-2g(\gamma ) & 0 & 0 \\
0 & 0 & 0 & 0 \\
\frac{\gamma }{2} & 0 & 0 & g(\gamma )%
\end{array}%
\right)
\end{equation}%
with%
\begin{equation*}
g(\gamma )=\left\{
\begin{array}{cc}
\frac{\gamma }{2} & \gamma \geq \frac{2}{3} \\
\frac{1}{3} & \gamma <\frac{2}{3}%
\end{array}%
\right. .
\end{equation*}%
The state is entangled for all nonzero $\gamma $ due to its concurrence [13]
$C(\varrho _{m})=\gamma $. The state was shown to violate the previous Bell
inequality only for $\gamma >0.8$. Consider one of its decompositions, the
state can be written by
\begin{eqnarray}
\varrho _{m} &=&\left[ g(\gamma )+\frac{\gamma }{2}\right] \left\vert \Phi
^{+}\right\rangle \left\langle \Phi ^{+}\right\vert +\left[ g(\gamma )-\frac{%
\gamma }{2}\right] \left\vert \Phi ^{-}\right\rangle \left\langle \Phi
^{-}\right\vert  \notag \\
&&+\left[ 1-2g(\gamma )\right] \left\vert 01\right\rangle \left\langle
10\right\vert ,
\end{eqnarray}%
where $\left\vert 01\right\rangle =\frac{1}{\sqrt{2}}\left( \left\vert \Psi
^{+}\right\rangle +\left\vert \Psi ^{-}\right\rangle \right) $ and $%
\left\vert \Phi ^{\pm }\right\rangle =\frac{1}{\sqrt{2}}\left( \left\vert
00\right\rangle \pm \left\vert 11\right\rangle \right) $ and $\left\vert
\Psi ^{\pm }\right\rangle =\frac{1}{\sqrt{2}}\left( \left\vert
01\right\rangle \pm \left\vert 10\right\rangle \right) $ are four Bell
states written in computational basis. Consider the correlation function $%
P(\theta _{1},\theta _{2})$ given by
\begin{equation}
P(\theta _{1},\theta _{2})=tr\varrho _{m}\left[ \mathcal{A}(\theta
_{1})\otimes \mathcal{B}(\theta _{2})\right]
\end{equation}%
where
\begin{equation*}
\mathcal{A}(\theta _{i})=\cos \theta _{i}\left( \left\vert 0\right\rangle
\left\langle 0\right\vert -\left\vert 1\right\rangle \left\langle
1\right\vert \right) +\sin \theta _{i}\left( \left\vert 0\right\rangle
\left\langle 1\right\vert +\left\vert 1\right\rangle \left\langle
0\right\vert \right)
\end{equation*}%
and $\mathcal{B}(\theta _{2})$ are defined analogously, and substitute the
decomposition given by eq. (10) associated with the corresponding
correlation functions into inequality (6), one can obtain the corresponding
Bell inequality. Numerical optimization to maximize the violation shows that
the inequality is violated for all $\gamma >0$. See Fig. 1. Note that
"maximize the violation" means maximizing $<B>=\left\vert \left\langle
ab\right\rangle -\left\langle ac\right\rangle \right\vert
+\sum_{i=1}^{N}p_{i}\left\vert \left\langle db\right\rangle
_{i}+\left\langle dc\right\rangle _{i}\right\vert $ in the paper.

\begin{figure}[tbp]
\includegraphics[width=8.5cm]{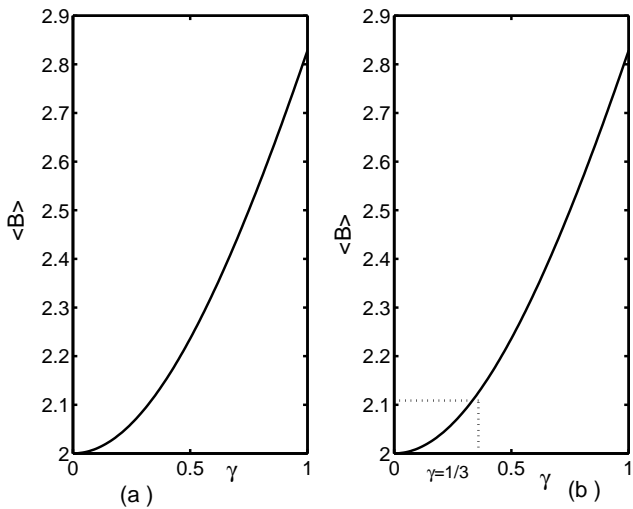}
\caption{Plot of the maximum violation of our Bell inequality versus $%
\protect\gamma$. (a) corresponds to the state $\protect\varrho_m$ which
shows the inequality can be violated for all $\protect\gamma>0$; (b)
corresponds to the Werner state, i.e. the state $\protect\varrho_w(\protect%
\gamma,\protect\pi/4)$, which shows that our inequality can be violated for
all $\protect\gamma>1/3$. (b) also shows the inequality can be violated for $%
\protect\gamma \leq 1/3$ due to the considered decomposition of $\protect%
\varrho_w$ }
\label{1}
\end{figure}

Another example is the variational Werner state introduced in Ref. [5] given
by

\begin{equation}
\varrho _{w}(\gamma,\xi )=\frac{1-\gamma }{4}I_{2}\otimes I_{2}+\gamma
\left\vert \Psi _{non}\right\rangle \left\langle \Psi _{non}\right\vert ,
\end{equation}%
where $I_{2}\ $is $(2\times 2)$-dimensional identity matrix and $\left\vert
\Psi _{non}\right\rangle =\cos \xi \left\vert 00\right\rangle +\sin \xi
\left\vert 11\right\rangle $. For $\xi =\frac{\pi }{4}$, eq. (12) is the
usual Werner state which was the first state found to be entangled for $%
\gamma >\frac{1}{3}$[11,14] and not violate a Bell inequality for single
states. The Werner state was shown to violate the Bell inequality in Ref.
[11] only for its concurrence $C(\varrho _{w})>\sqrt{\frac{1}{3}}$. Consider
a possible decomposition as%
\begin{eqnarray*}
\varrho _{w}(\gamma ,\pi/4 ) &=&\frac{1-\gamma }{4}\left( \left\vert \Phi
^{+}\right\rangle \left\langle \Phi ^{+}\right\vert +\left\vert \Psi
^{+}\right\rangle \left\langle \Psi ^{+}\right\vert \right. \\
&&\left. +\left\vert \Phi ^{-}\right\rangle \left\langle \Phi
^{-}\right\vert +\left\vert \Psi ^{-}\right\rangle \left\langle \Psi
^{-}\right\vert +\gamma \left\vert \Psi _{non}\right\rangle \left\langle
\Psi _{non}\right\vert ,\right)
\end{eqnarray*}%
and the analogous correlation function given by eq. (11), one can obtain the
corresponding Bell inequality. By optimization to maximize the violation
(see Fig. 2), one can find that the state $\varrho _{w}(\gamma )$ violates
the Bell inequality for all $\gamma >\frac{1}{3}.$

\begin{figure}[tbp]
\includegraphics[width=8.5cm]{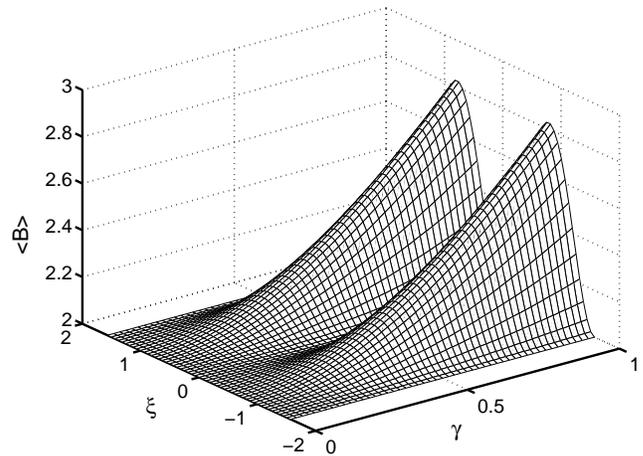}
\caption{The maximum violation of our inequality for the variational
Werner state $\protect\varrho_w{(\protect\gamma,\protect\xi)}$
versus $\protect\gamma$ and $\protect\xi$. The figure shows the
periodic violation of the state with $\protect\xi$ and the violation
with $\protect\gamma$. } \label{2}
\end{figure}

Above examples have shown that they violate our inequality by considering
proper decompositions, although the original CHSH inequality is not
violated. In our opinion, the key lies in the constraint on the Bell
operator, $\left\langle ab\right\rangle -\left\langle ac\right\rangle $.
I.e. the bound on the Bell operator in the original CHSH inequality is not
tight enough for any entangled mixed state. Ours can be regarded as a
correction of the bound. In this sense, we say our inequality is more
refined.

What's more, from Fig. 2 and Fig. 1 (b), it is so surprising that the
inequality is violated not only for $\gamma >\frac{1}{3}$ but for all $%
\gamma >0$, which means a separable mixed state can also violate the
inequality. It seems to be a paradox. In fact, it is not the case. The key
lies in that our inequality depends on the decomposition of the density
matrix. To better show the dependent relation, let us take a third density
matrix as an example. Consider the bipartite density matrix given by
\begin{equation}
\varrho _{s}=\left(
\begin{array}{cccc}
\frac{1}{4} & 0 & 0 & x \\
0 & \frac{1}{4} & x & 0 \\
0 & x & \frac{1}{4} & 0 \\
x & 0 & 0 & \frac{1}{4}%
\end{array}%
\right) ,
\end{equation}%
with $\left\vert x\right\vert \leq \frac{1}{4}$, one can have $C(\varrho
_{s})=0$ for all $\left\vert x\right\vert \leq \frac{1}{4}$. That is to say,
$\varrho _{s}$ can expressed by the convex combination of product states,
i.e.
\begin{eqnarray}
\varrho _{1} &=&\varrho _{s}=\left( \frac{1}{4}+x\right) \left\vert \varphi
\right\rangle \left\langle \varphi \right\vert \otimes \left\vert \varphi
\right\rangle \left\langle \varphi \right\vert  \notag \\
&&+\left( \frac{1}{4}-x\right) \left\vert \varphi \right\rangle \left\langle
\varphi \right\vert \otimes \left\vert \psi \right\rangle \left\langle \psi
\right\vert  \notag \\
&&+\left( \frac{1}{4}-x\right) \left\vert \psi \right\rangle \left\langle
\psi \right\vert \otimes \left\vert \varphi \right\rangle \left\langle
\varphi \right\vert  \notag \\
&&+\left( \frac{1}{4}+x\right) \left\vert \psi \right\rangle \left\langle
\psi \right\vert \otimes \left\vert \psi \right\rangle \left\langle \psi
\right\vert ,
\end{eqnarray}%
where $\left\vert \varphi \right\rangle =\frac{1}{\sqrt{2}}\left( \left\vert
0\right\rangle +\left\vert 1\right\rangle \right) $ and $\left\vert \psi
\right\rangle =\frac{1}{\sqrt{2}}\left( \left\vert 0\right\rangle
-\left\vert 1\right\rangle \right) $. However, $\varrho $ can also obtained
by the convex combination of maximally entangled states, i.e. \bigskip
\begin{eqnarray}
\varrho _{2} &=&\varrho _{s}=\left( \frac{1}{4}+x\right) \left( \left\vert
\Phi ^{+}\right\rangle \left\langle \Phi ^{+}\right\vert +\left\vert \Psi
^{+}\right\rangle \left\langle \Psi ^{+}\right\vert \right)  \notag \\
&&+\left( \frac{1}{4}-x\right) \left( \left\vert \Phi ^{-}\right\rangle
\left\langle \Phi ^{-}\right\vert +\left\vert \Psi ^{-}\right\rangle
\left\langle \Psi ^{-}\right\vert \right) .
\end{eqnarray}%
Considering the same correlation functions and following the same procedure,
based on the inequality (6), one can obtain the corresponding Bell
inequality for eq. (14) and eq. (15), respectively. By our numerical
optimization to maximize the violation of the inequalities for eq. (14) and
eq. (15), respectively, given by Fig. 3, one can find that $\varrho _{2}$
always violate the inequality for $nozero$ $x$, while $\varrho _{1}$ is
always constrained by the inequality for all $x$. This just shows the
property that the current inequality depends on the decomposition of density
matrix. In fact, if keeping it in mind that all pure states cannot violate
the original CHSH inequality, one will easily find from the derivation of
our inequality that a separable density matrix cannot violate our inequality
if considering the product-state-decomposition.
\begin{figure}[tbp]
\includegraphics[width=8.5cm]{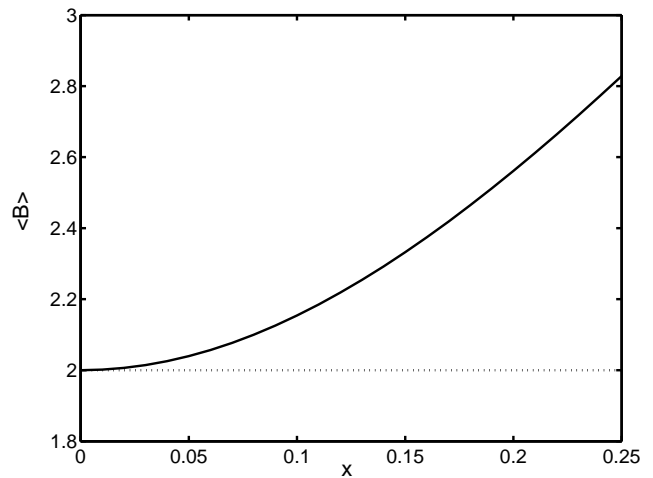}
\caption{The maximum violation of our inequality for the separable state $%
\protect\varrho _{s}$ versus $x$ in terms of two different decompositions.
The figure shows that the inequality can be violated for the entangled-state
decomposition (solid line) and can not be violated for the product-state
decomposition (dotted line).}
\label{3}
\end{figure}

Since the violation of Bell inequality means there exists quantum
correlation, our examples have shown that a separable mixed state
may have quantum correlation which depends on the concrete
realization of the state, even though the state has been defined as
a separable one based on the usual entanglement measure such as
concurrence and so on [17]. In fact, this is not strange. As
mentioned in Ref. [5], the classical correlation does not mean the
state has been prepared in the manner described, but only that its
statistical properties can be reproduced by a classical mechanism.
In other words, if considering the entanglement of pure states as a
cost, the usual measurement of entanglement of formation for mixed
states just gives the least cost to reproduce the mixed states. That
is to say, the usual entanglement measure does not always extract
quantum correlations that have been used to generate the given mixed
state. I.e. The violation of our inequality means that quantum
correlations are needed to produce the given mixed state by the
considered concrete realization (decomposition). In this sense, we
say that a separable mixed state may \textit{owe} some quantum
correlations. Therefore, in order to demonstrate whether a mixed
state owe quantum correlations in terms of previous entanglement
measures or whether our inequality is consistent with the usual
entanglement measures, one has to test whether our inequality is
violated in terms of the optimal decomposition in the sense of the
given entanglement measure (for example, concurrence and so on).

In summary, we have presented a generalized Bell inequality. The inequality
has been shown to be more refined than the previous ones. The most important
property is that the inequality has a variable bound which depends on the
decomposition of the state. As a result, a separable quantum mixed state may
be shown to include quantum correlation, a potential understanding of which
has been provided. Finally, we hope that the current result will further the
understanding of quantum entanglement and quantum nonlocality.

Thank X. X. Yi, C. Li and Y. Q. Guo for their valuable discussions. This
work was supported by the National Natural Science Foundation of China,
under Grant Nos. 10575017 and 60472017.


\begin{thebibliography}{99}
\bibitem{[1]} A. Einstein, B. Podolsky, and N. Rosen, Phys. Rev. \textbf{47}%
, 777 (1935).

\bibitem{[2]} S. L. Braunstein, A. Mann, and M. Revzen, Phys. Rev. Lett.
\textbf{68}, 3259 (1992).

\bibitem{[3]} J. S. Bell, Physics (Long Island City, N.Y.) \textbf{1}, 195
(1964).

\bibitem{[4]} J. Clauser, M. Horne, A. Shimony, and R. Holt, Phys. Rev.
Lett. \textbf{23}, 880 (1969).

\bibitem{[5]} R. F. Werner, Phys. Rev. A \textbf{40}, 4277 (1989).

\bibitem{[6]} K. Banaszek and K. W\'{o}dkiewicz, Phys. Rev. A \textbf{58},
4345 (1998); Phys. Rev. Lett. \textbf{82}, 2009 (1999); Acta Phys. Slovaca
\textbf{49}, 491 (1999).

\bibitem{[7]} H. Jeong, J. Lee, and M. S. Kim, Phys. Rev. A \textbf{61},
052101 (2000).

\bibitem{[8]} Zeng-Bing Chen, Jian-Wei Pan, Guang Hou and Yong-De Zhang,
Phys. Rev. Lett. \textbf{88}, 040406 (2002).

\bibitem{[9]} N. Gisin, Phys. Lett. A \textbf{145}, 201 (1991).

\bibitem{[10]} S. Popescu and D. Rohrlich, Phys. Lett. A \textbf{166}, 293
(1992).

\bibitem{[11]} N. Gisin and A. Peres, Phys. Lett. A \textbf{162}, 15 (1992).

\bibitem{[12]} S. L. Braunstein, A. Mann, and M. Revzen, J. Phys. A \textbf{%
25}, L851 (1992).

\bibitem{[13]} W. J. Munro, K. Nemoto and A. G. White, J. Mod. Opt. \textbf{%
48}(7), 1239 (2001).

\bibitem{[14]} A. G. White, D. V. F. James, W. J. Munro, and P. G. Kwiat,
(submitted to Nature).

\bibitem{[15]} C. H. Bennett, G. Brassard, S. Popescu, B. Schumacher, and W.
K. Wootters, Phys. Rev. Lett. \textbf{76}, 722 (1996).

\bibitem{[16]} An alternate derivation can be obtained by considering the
convex combination of the CHSH inequalities which each pure state of the
density matrix satisfies and utilizing the absolute value inequality.

\bibitem{[17]} W. K. Wootters, Phys. Rev. Lett. \textbf{80}, 2245 (1998);
and the references therein.
\end{thebibliography}
\end{document}